\documentclass[twocolumn,secnumarabic,amssymb, nobibnotes, aps, prd]{revtex4-1}

\usepackage{graphicx}
\usepackage{amsmath}

\begin{document}

\title{Theory of Landau level mixing in heavily graded graphene p-n junctions}%

\author{Samuel W LaGasse}%
\email[Samuel W LaGasse: ]{slagasse@sunypoly.edu}
\affiliation{SUNY Polytechnic Institute}
\author{Ji Ung Lee}%
\email[Ji Ung Lee: ]{jlee1@sunypoly.edu}
\affiliation{SUNY Polytechnic Institute}
\date{August 2, 2016}%

\begin{abstract}
We demonstrate the use of a quantum transport model to study heavily graded graphene p-n junctions in the quantum Hall regime. A combination of p-n interface roughness and delta function disorder potential allows us to compare experimental results on different devices from the literature. We find that wide p-n junctions suppress mixing of $n \neq 0$ Landau levels. Our simulations spatially resolve carrier transport in the device, for the first time, revealing separation of higher order Landau levels in strongly graded junctions, which suppresses mixing.
\end{abstract}

\maketitle

\section{Introduction}

The discovery of the integer quantum Hall effect in 1980 was a seminal event in the field of condensed matter physics \cite{Klitzing1980}. Shortly thereafter, the fractional quantum Hall effect was also discovered \cite{Tsui1982}. The observation of integer and fractional steps in the Hall conductance, in units of $e^2/h$, cemented two dimensional electron gases (2DEGs) as a platform to study quantum transport. Conventional 2DEGs formed in semiconductor heterostructures, however, are restricted to unipolar conduction, either by electrons or holes. The discovery of graphene in 2004 by Novoselov and Geim lifted this restriction, giving physicists a fascinating material to investigate the quantum Hall effect in devices with ambipolar conduction \cite{Williams2007}. 

%%%%%%%%%%%%%%%%%%%%%%%%%%%%%%%%%%%
%%%%%%%      Schematic      %%%%%%%
%%%%%%%%%%%%%%%%%%%%%%%%%%%%%%%%%%%
\begin{figure}
\includegraphics[width=0.5\textwidth]{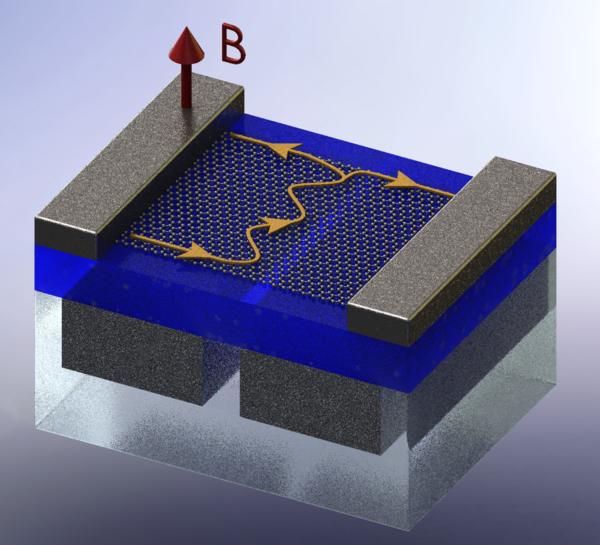}
\caption{A schematic of a graphene device with buried split-gates. The device consists of a silicon dioxide substrate, into which two metal gates are formed. A gate oxide on top of the silicon and metal gates electrically isolates the graphene from the gates. Contacts on each side of the graphene device allow for measurements of carrier transport. The split-gates capacitively couple to the graphene sheet and are modulate the local Fermi level. By varying the voltage of the two gates separately, it is possible to form different types of junctions, including a p-n junction. Applying a strong external magnetic field, perpendicular to the graphene sheet, causes the carriers to be constricted into Landau levels, which travel around the edges of the graphene. The arrows depict the flow of carriers in the lowest Landau level.}
\label{fig:schematic}
%%%%%%%%%%%%%%%%%%%%%%%%%%%%%%%%%%%
%%%%%%%      Schematic      %%%%%%%
%%%%%%%%%%%%%%%%%%%%%%%%%%%%%%%%%%%
\end{figure}

Graphene is formed by carbon atoms arranged in a single layer honeycomb structure, yielding a gap-less band structure with linear Dirac cones at two degenerate $\textbf{K}$ and $\textbf{K}'$ points \cite{Novoselov2005}. At the $\textbf{K}$ and $\textbf{K}'$ points the two-fold spin degeneracy is split between electron and hole carriers, yielding a characteristic half-integer quantum Hall effect \cite{Novoselov2005}\cite{Zhang2005}. In addition to ambipolar conduction, graphene exhibits extremely high carrier mobility \cite{Kim2009} and thermal conductivity \cite{Balandin2008}; these properties make it a candidate channel material for future electronic applications \cite{Schwierz2010}.  

In this paper, we present a model studying the transition between graphene p-n junctions which mix Landau levels \cite{Williams2007} and those which only mix the lowest Landau level \cite{NikolaiN.Klimov2015a}. We seek to further understand the magnetotransport of p-n junctions formed with buried split-gates, as depicted in Fig. (\ref{fig:schematic}). In order to study the effect of junction width, $D_W$, on transport, we combine the delta function disorder model of \cite{Li2008} and p-n junction interface roughness model of \cite{Low2009a}, giving a simulation with more realistic conditions. We will start by introducing the details of the model, demonstrate how it can be used to replicate experimental results of \cite{Williams2007} and \cite{NikolaiN.Klimov2015a}, and then present several visualizations which assist in understanding the underlying transport mechanisms.
\section{Background}
Through the use of metal gates capacitively coupled to graphene, it is possible to create a p-n junction by locally modulating the carrier concentration. The p-n junction is a fundamental device, used as the building block from which many other devices are built. In graphene, p-n junctions exhibit very interesting physics such as Klein tunneling\cite{Katsnelson2006} and may be used in electron optics \cite{Cheianov2007}. 

The application of a magnetic field perpendicular to a graphene device produces a Lorentz force which constricts transporting carriers to the edges of the sheet. A sufficiently strong magnetic field will confine the carriers into edge states, known as Landau levels, whose energy is given by 
\begin{equation}
E_{LLn} = \textrm{sgn}(n) \ \sqrt[]{\mid n \mid 2eB\hbar\nu^2_F}
\label{energy_levels}
\end{equation}
where $n$ is the Landau level index, given by an integer. The term $e$ is the electron charge, $B$ is the applied magnetic field, $\hbar$ is the reduced Planck's constant, and $\nu_F$ is the Fermi velocity (approximately $10^6 \ \textrm{cm/s}$). 

In a typical graphene quantum Hall measurement, the entire graphene device is uniformly doped by a global back gate and a strong magnetic field is applied. When the Fermi energy is not set to the energy of the Landau level, $E_F \neq E_{LLn}$, the edge states on the opposite sides of the channel are isolated by an insulating bulk state. In this configuration, carrier conduction only takes place through the edges of the device. When the back gate voltage is modulated and the Fermi energy moves through the energy of a Landau level, $E_F \approx E_{LLn}$, the bulk of the device no longer isolates the edges and electrons conduct through the entire device. These two conditions result in the transverse and longitudinal resistance, respectively, typically reported in experiments.

When a graphene p-n junction is formed, in the quantum Hall regime, the device simultaneously conducts through the edge states and localized bulk Landau levels. Away from the junction, when $E_F \neq E_{LLn}$, carriers conduct along the edge as before. However, at the junction, where the potential of the device smoothly transitions between n and p type, there will exist an equipotential line for each transporting edge state where \begin{equation}
E_F - E_{LLn} = E_{\textrm{on-site}}(x).
\label{ELL_condition}
\end{equation}
The term $E_{\textrm{on-site}}$ is the local potential energy in the device. On this equipotential line, carriers will conduct through the bulk, bridging the edge states on the opposite sides of the channel. Furthermore, in a very smooth p-n junction, the equipotential line of each bulk state will separate, allowing one to spatially resolve conduction in each Landau level. In this work we will use quantum transport calculations to verify the condition in (\ref{ELL_condition}) and spatially resolve conduction through the bulk in a smoothly graded p-n junction. 
%In an ambipolar graphene p-n junction, the carrier type switches from holes to electrons at the p-n interface. For a device in a magnetic field, the flipped sign of the carrier charge will also flip the direction of the Lorentz force, resulting in what is known as a snake state (depicted with arrows in Fig. (\ref{fig:schematic})). Snake states bridge the two edges of the graphene sheet, allowing carriers to switch sides in the device.

Abanin \textit{et al} predicted that when the Landau levels in a graphene p-n junction mix, plateaus in the two-terminal conductance will occur according to
\begin{equation}
G_{\textrm{two-terminal}} = \frac{\left| \nu_1 \right|\left| \nu_2 \right| }{\left| \nu_1 \right|+ \left| \nu_2 \right|}
\label{conductance_theory}
\end{equation}
where $\nu_{1,2} = [\pm2,\pm6,\pm10, ...]$ are the filling factors of the left/right sides of the junction. This effect was experimentally measured by Williams \textit{et al}, where several of the predicted plateaus were observed \cite{Williams2007}. The device  of Williams \textit{et al} was fabricated with a global back gate and local top gate, which were used to create the junction \cite{Williams2007}.

There have been several studies which model the results observed by Williams \textit{et al}. Tworzyd\l o \textit{et al} analyzed the importance of the valley-isospin and intervalley scattering \cite{Tworzydlo2007}. Long \textit{et al} \cite{Long2008a} and Li \textit{et al} \cite{Li2008} both demonstrated a quantum transport model including large on-site disorder delta function potentials which allowed the Landau levels to mix and demonstrated plateaus in unipolar and ambipolar junctions. Low \cite{Low2009a} presented an alternative quantum transport model which used interface roughness, edge roughness, and localized scattering centers to mix the Landau levels, tying closely to the experiments by Williams \textit{et al} \cite{Williams2007}.

Recently, Klimov \textit{et al} performed measurements on a graphene p-n junction which, in the ambipolar regime, only showed one plateau with a conductance of $1 \ e^2/h$ \cite{NikolaiN.Klimov2015a}. This single plateau was predicted to occur when only the $0^{\textrm{th}}$ Landau level mixes. The device of Klimov \textit{et al} was formed using a pair of split-gates buried 100 nm under the gate oxide, with a large inter-gate spacing. These split-gates produce a very graded junction profile, on the order of several hundred nanometers. The authors posited that the graded junction would spatially separate the higher order Landau levels, inhibiting mixing \cite{NikolaiN.Klimov2015a}. In contrast, the top gate used by Williams \textit{et al} is located very close to the graphene, producing a sharper junction \cite{Williams2007}.

\section{Modeling techniques}
%%%%%%%%%%%%%%%%%%%%%%%%%%%%%%%%%%%
%%%%%%% Benchmarking figure %%%%%%%
%%%%%%%%%%%%%%%%%%%%%%%%%%%%%%%%%%%
\begin{figure*}
\includegraphics[width=\textwidth]{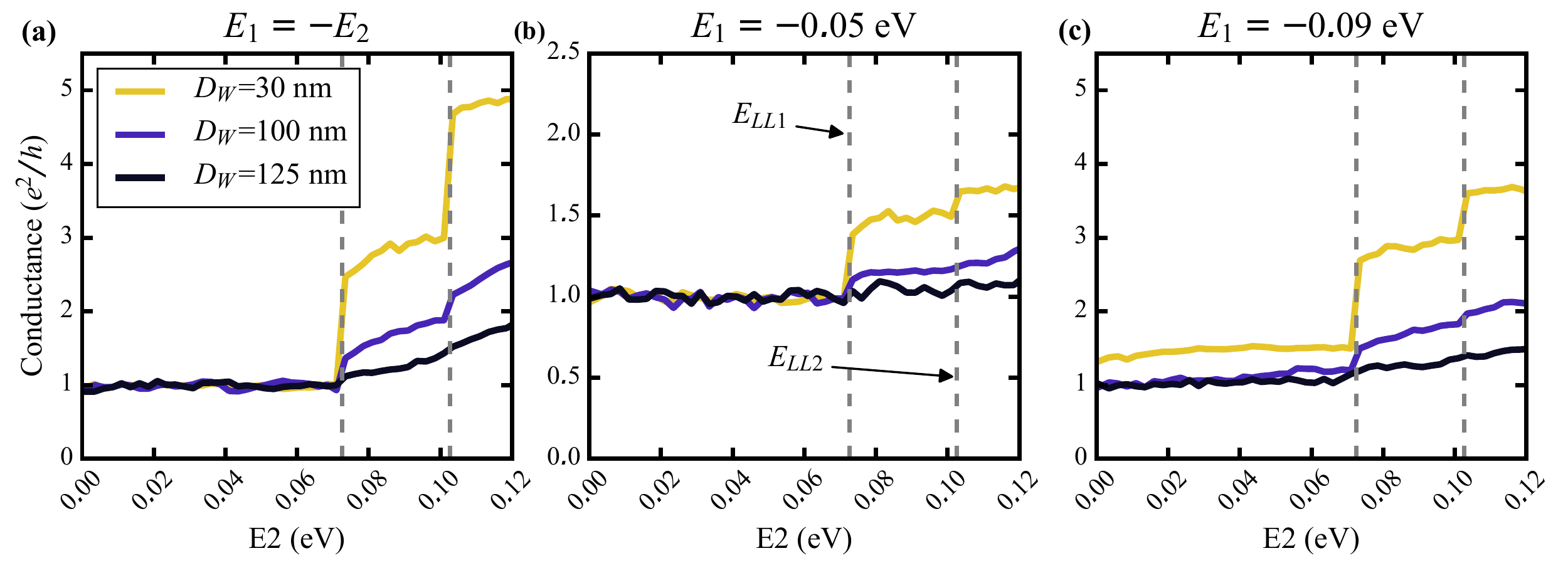}
\caption{We benchmark our model by simulating conductance between source and drain contacts, which is equivalent to Hall conductance measured in typical experiments. We compare our simulated conductance with experiments which show several plateaus in ambipolar conduction \cite{Williams2007} and those which do not \cite{NikolaiN.Klimov2015a}. All curves are for L = 320 nm, W = 200 nm, and B = 4 T. Each point in every curve is the ensemble average of the conductance for 400 different disorder realizations. \textbf{(a)} The diagonal slice of the conductance map typically measured. We set the on-site energy of the left side of the junction, $E_1$, to be the negative of the right side, i.e. $E_1 = -E_2$. This produces a symmetric n-p junction. \textbf{(b)} The on-site energy of the left side of the junction is fixed to $E_1 = -0.05$ eV, and the right side on-site energy is varied.\textbf{(c)} The junction is configured as in \textbf{(b)}, but now $E_1 = -0.09$ eV. In each plot, when the junction width is small ($D_W=30$ nm), all the Landau levels are mixed. Plateaus occur when the filling factors jump from 2 to 6 to 10 (the transitions are indicated by vertical dashed lines). When the junction widths are longer, mixing is suppressed and only the lowest Landau level fully mixes. This is consistent with experimental results in the literature.}
\label{transmission}
\end{figure*}
%%%%%%%%%%%%%%%%%%%%%%%%%%%%%%%%%%%
%%%%%%% Benchmarking figure %%%%%%%
%%%%%%%%%%%%%%%%%%%%%%%%%%%%%%%%%%%
In the past, quantum transport modeling has been used ubiquitously to great success in capturing the physics of graphene transport both without \cite{Low2009,Sajjad2012,Sajjad2013a} and with magnetic fields present\cite{Li2008,Low2009a,Long2008a,Golizadeh-mojarad2006}. In this work, we will use the scattering matrix (S-matrix) method, which enables us to calculate the terminal characteristics of the device in Fig. (\ref{fig:schematic}). Using the S-matrix method, we are also able to calculate the wave-function inside the device channel, allowing for visualizations of carrier transport. The numerical aspects of the calculations were performed using the quantum transport package KWANT \cite{Groth2014}. 

Carrier transport at low energies in graphene is described by a massless Dirac Hamiltonian given by
\begin{equation}
\hat{H} = \nu_F \boldsymbol{\sigma} \cdot \mathbf{p}
\end{equation}
where $\boldsymbol{\sigma} = \left(\sigma_x,\sigma_y\right)$ is a vector of Pauli matrices and $\textbf{p} = \left(\hat{p}_x, \hat{p}_y\right)^\textrm{T}$ is a vector of momentum operators. This Hamiltonian may be discretized onto a honeycomb lattice, resulting in the tight-binding Hamiltonian
\begin{equation}\label{Hamiltonian}
\hat{H}= \sum_{i}^N\epsilon_i\hat{c}_i^\dag\hat{c}_i+\sum_{i,j}^Nt_{i,j}\hat{c}_i^\dag\hat{c}_j,
\end{equation}
written in the language of creation/annilation operators $\hat{c}_i^\dag / \hat{c}_j$. 

The first summation in (\ref{Hamiltonian}) fills the Hamiltonian matrix diagonals with $\epsilon_i$, the on-site energy at site $i$. The on-site energy describes the potential landscape of the device and allows the creation of a p-n junction. The second summation in (\ref{Hamiltonian}) only generates non-zero matrix elements for lattice sites which are first nearest-neighbors, allowing transport between the sites. 

Typically, $t_{i,j}$ is set to $t_0 = 2.71 \textrm{ eV}$, representing the $\pi-$bond overlap between first nearest-neighbor atoms \cite{Reich2002}. In this work we adopt a scaled tight-binding model, first presented in \cite{Liu2015}, where the lattice constant of graphene $a_0$ and the hopping parameter $t_0$ are scaled by a scaling factor $s_f$ according to
\begin{equation}
\begin{split}
a=s_{f}a_{0} \\ t_{i,j}=t_{0}/s_f.
\end{split}
\end{equation}
We use a scaling factor of $s_f=8$, which allows for more efficient simulations whilst still accurately capturing the physics of graphene.

In order to include the effect of a magnetic field applied perpendicular to the graphene sheet, we introduce Peierl's phase by setting $\textbf{p} \longrightarrow \textbf{p}-e\textbf{A}$. \textbf{A} is the magnetic vector potential and the magnetic field is given by $\textbf{B}=\nabla \times \textbf{A}$. For a device with leads oriented along the x-direction, it is convenient to define the magnetic vector potential using Landau gauge; $\textbf{A}=\langle 0,-Bx,0 \rangle$ where $B$ is the magnitude of the applied magnetic field. The effect of Peierl's phase, in this case, is to modify the hopping parameter according to 
\begin{equation}\label{peierls}
t_{i,j} = t\exp\left[i\frac{e}{\hbar}\int^{\textbf{r}_j}_{\textbf{r}_i} \textbf{A}\cdot\textbf{dr}\right],
\end{equation}
where $t$ is the unperturbed hopping parameter. The integral in (\ref{peierls}) is a line integral which takes place between the two sites $i$ and $j$ and may be calculated as a straight line, yielding
\begin{equation}\label{simplified_peierls}
t_{i,j} = t \exp\left[-i\frac{e}{\hbar}	B\left( x_i-x_j\right)\frac{y_i+y_j}{2}\right].
\end{equation}

Simply simulating a pristine abrupt p-n junction in graphene, by putting a step in the on-site energy profile, is not sufficient to capture experimentally observed quantum Hall effects. It is necessary to include some extrinsic effects to mix Landau levels and cause inter-valley scattering at the p-n junction interface. In addition, experimentally realized p-n junctions have a finite transition between the n and p regions, the junction width $D_W$, which must be included. In this paper, we present a model combining the interface delta function disorder of \cite{Li2008}, the p-n interface roughness model of \cite{Low2009a}, and a finite $D_W$. Our model uses disorder potential a factor of four less than that used in \cite{Li2008}. The lower disorder potential is needed to demarcate the junction profile and combined with the roughness model of \cite{Low2009a}, allows us to study junction width effects in experimentally measured p-n junctions.

In our simulations, we perform an ensemble average of conductance over many randomly generated disorder profiles for a device with a fixed interface roughness profile. This procedure is used to account for ergodicity, which states that time averaging by measurements in the lab may be accounted for in simulations by ensemble averages of systems with spatial disorder \cite{Beenakker1997a}. 

The p-n junction profile, including effects of junction width, interface disorder, and interface roughness, are all included by modifying the on-site energy in the device channel Hamiltonian (\ref{Hamiltonian}). We implement modifications to the on-site energy according to the piecewise function
\[ \begin{cases} 
      E_1, & x\leq I(y)-D_W/2 \\
      \frac{E_2-E_1}{D_W}x+\frac{E_1+E_2}{2}+\delta_i, & \parbox[t]{.4\columnwidth}{$I(y)-D_W/2\leq x\leq \qquad\qquad I(y)+D_W/2$} \\
      E_2, & x\geq I(y)+D_W/2 
   \end{cases}
\] 

$E_{1,2}$ represent the shift in the on-site energy in the device produced by capactively coupled gates. In the context of this work, a positive shift in $E_{1,2}$ creates a p-type region and a negative shift in $E_{1,2}$ creates an n-type region. The term $\delta_i$ is a delta function disorder potential placed each site, $i$,in the junction transition region. 

The delta function disorder term, $\delta_i$, added to the on-site energy at the sites in the junction transition region, is randomly generated according to a Gaussian distribution centered at $0.0 \textrm{ eV}$ with a standard deviation of  $0.15 \textrm{ eV}$. This site-to-site change in potential energy is sufficient to cause intervalley scattering at the junction, which is necessary to capture experimental results. The maximum disorder potential in our model is a factor of four smaller than that suggested by Li \textit{et al} \cite{Li2008}. Using such a large disorder potential would obscure the effect of junction width on Landau level mixing. In our case, the disorder potential perturbs the Landau levels, but the effect of junction width will still be seen. 

$I(y)$ is the junction interface roughness profile, created using the model presented by Low\cite{Low2009a}. We will repeat the specifics of the interface roughness model here for clarity. $I(y)$ is generated as a Fourier series, given by 
\begin{equation}
\label{roughness_profile}
I(y)=\sum_{n}^N A_n \sin\left(\frac{n\pi y}{W}\right).
\end{equation}
The amplitude of the $n^{\textrm{th}}$ Fourier component is defined as
\begin{equation}
A_n=R(D_1) e^{-\frac{n}{D_2}}.
\end{equation}
The function $R(D_1)$ gives a uniformly distributed random number around $\pm D_1$. The terms $D_{1,2}$ and the number of Fourier components, $N$, are used to control the form of the roughness profile. In our simulations, we set $D_{1,2}=13$ and $N=30$. This yields a roughness profile with an RMS standard deviation of approximately 12 nm.

Now that the device Hamiltonian (\ref{Hamiltonian}) is fully defined, we use the S-matrix formalism to study its transport properties. We calculate the zero temperature, zero bias two-terminal conductance according to the equation
\begin{equation}
 G(E) = \frac{2e^2}{h}T(E),
 \end{equation} 
where the two is for spin degeneracy. The zero temperature, zero bias approximation is valid when comparing to low bias measurements performed at, or below, liquid helium temperatures. $T(E)$ is the quantum mechanical transmission function given by
\begin{equation}
T(E)=\sum_{n\in S, m\in D} \mid S_{nm}(E)\mid ^2,
\end{equation}
where the summation occurs over the S-matrix elements connecting the source and drain contacts to the channel. The calculation of the transmission function automatically includes another factor of two for valley degeneracy, which is intrinsically factored into our Hamiltonian. 

In addition to calculating the conductance, we also calculate the wave functions associated with the Landau levels. By spatially resolving the wave functions, we produce maps of the transporting electron probability density, which are useful for analyzing the underlying physical mechanisms of conduction for each Landau level. The probability density at site $i$ is given by 
\begin{equation}
\rho(i) = \sum_{j = S,D} \psi_j(i)\psi_j(i)^*,
\label{wavefunc}
\end{equation}
where $\psi_j(i)$ is the wave function at site $i$ for carriers from the $j^\textrm{th}$ contact. In practice, it is helpful to remove the summation in (\ref{wavefunc}) and study the carriers injected by only one contact at a time. 
\section{Results and Discussion}
We will begin by comparing the results of our model with experimental results in the literature for ambipolar junctions which mix several Landau levels \cite{Williams2007} and those which only mix the lowest Landau level \cite{NikolaiN.Klimov2015a}. After verifying that our model is able to replicate results for different experimental junctions, we will seek to further understand the mechanisms at play. We will next study the effect of junction width and magnetic field strength on the degree of Landau level mixing in a junction. Several visualizations will be presented for pristine and disordered junctions which demonstrate the effect of a large junction width, where the Landau levels are separated and may be spatially resolved. Finally, we will compare a very wide junction device with analytical wave function calculations \cite{Lukose2007}, demonstrating how the graded p-n junction may reveal the effect of graphene's two sub-lattice structure. . 
\subsection{Comparison with experiment}
In numerically studying quantum transport, it is very important to first benchmark the model against some experimental measurements. We choose to model the experimental Hall conductance measurements of \cite{Williams2007} and \cite{NikolaiN.Klimov2015a}. In an experiment, one can sweep two gate voltages independently, measuring the Hall conductance at each gate voltage. This may be continued to generate a four quadrant map of conductance for the p-n, n-p, p-p, and n-n configurations. We will focus on the ambipolar junction configuration, which is what makes graphene special compared to conventional 2DEGs.

In Fig. (\ref{transmission}) we show the simulated Hall conductance for a diagonal slice and two horizontal slices (at $\nu_n=6 \textrm{ and }10$) of the n-p quadrant for $D_W = 30, 100, \textrm{ and } 125 \textrm{ nm}$. The x-axis of each plot shows the on-site shift of the right side of the junction, $E_2$. This shift in the on-site energy represents the effect of a charged gate nearby the graphene sheet. 

Each curve in Fig. (\ref{transmission}) is the ensemble average of 400 different randomly generated disorder configurations. There is an applied magnetic field of 4 T perpendicular to the graphene sheet. The device scattering region is 200 nm wide and 320 nm long. In this case, we choose a sheet with zigzag edges, but with our disorder model, armchair edges would yield very similar results. 

When the junction width is 30 nm, our simulation recovers the first three plateaus predicted by (\ref{conductance_theory}), which were first measured in \cite{Williams2007}. The plateaus occur at the energy levels predicted by (\ref{energy_levels}), denoted by vertical lines in Fig. (\ref{transmission}). At a junction width of 100 nm, we observe partial mixing of the first and second Landau levels. The plateaus are still visible, but occur at smaller values of conductance. 

As we increase the junction width to 125 nm, there is a slight increase in conductance for filling factors of 6 and 10 in the diagonal slice, but plateaus no longer form. For the horizontal slices at $\nu_n=6\textrm{ and }10$, the conductance is nearly flat at 1 $e^2/h$. This result is consistent with what was measured by Klimov \textit{et al}\cite{NikolaiN.Klimov2015a}.

\subsection{$D_W$ and magnetic field dependence}
%%%%%%%%%%%%%%%%%%%%%%%%%%%%%%%%%%%
%%%%%%%      DW Depend      %%%%%%%
%%%%%%%%%%%%%%%%%%%%%%%%%%%%%%%%%%%
\begin{figure}
\includegraphics[width=0.5\textwidth]{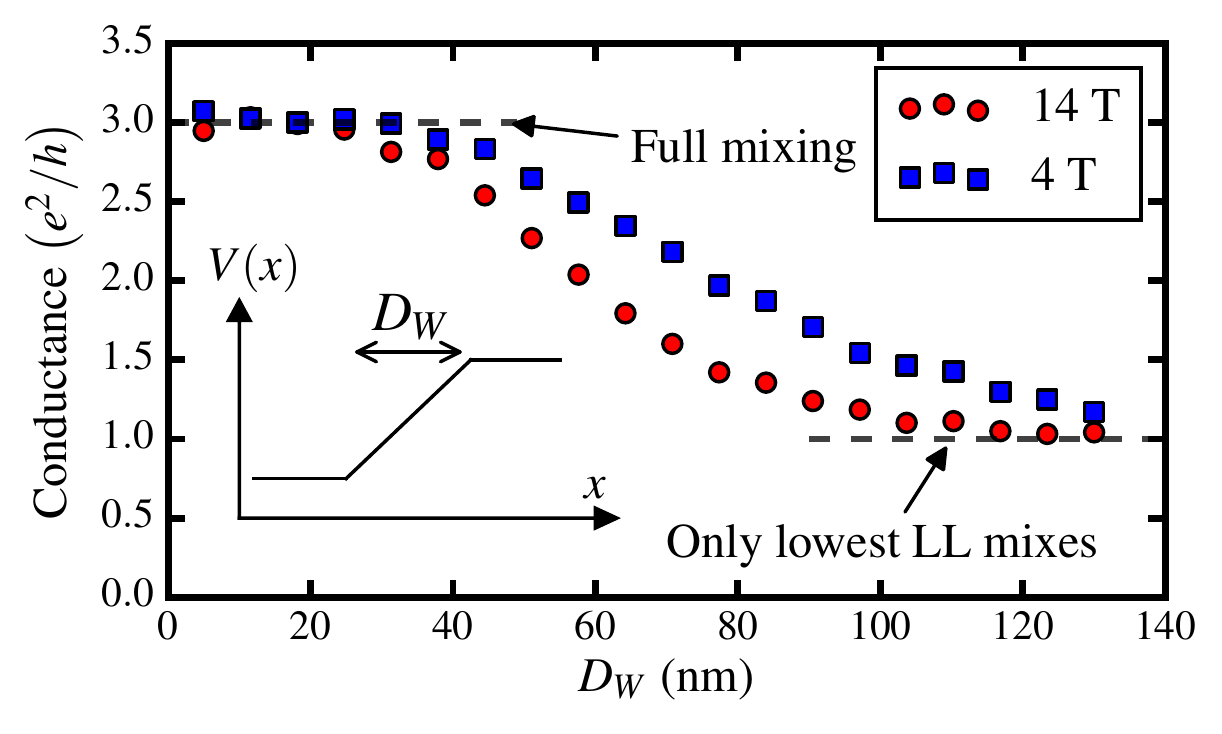}
\caption{Hall conductance as a function of junction width, $D_W$. The device is configured in the same way as Fig. (\ref{transmission}), but instead of varying the on-site energies in the device, we fix the filling factors. In this case, the device is configured as $E_1=-E_2=0.0875$ eV, or $\nu_n=-\nu_p=6$. The effect is demonstrated for two different magnetic fields: $B$ = 4 and 14 Tesla. When the junction fully mixes all of the Landau levels, for junction widths shorter than about 40 nm, the conductance is approximately 3.0 $e^2/h$. For junction widths over 100 nm, the device only mixes the lowest Landau level, yielding a conductance of approximately 1.0 $e^2/h$. Our simulations indicate that stronger magnetic fields cause junction width to have a more pronounced effect in restricting mixing of higher order Landau levels.}
\label{dw_dependence}
\end{figure}
%%%%%%%%%%%%%%%%%%%%%%%%%%%%%%%%%%%
%%%%%%%      DW Depend      %%%%%%%
%%%%%%%%%%%%%%%%%%%%%%%%%%%%%%%%%%%

Now that we have demonstrated that our model is able to capture the experimental Hall conductance of \cite{Williams2007} and \cite{NikolaiN.Klimov2015a}, we will seek to explain the differences between the two. The device which showed several plateaus \cite{Williams2007} was fabricated using a global back gate and local top gate to form a p-n junction. The device which showed a single plateau \cite{NikolaiN.Klimov2015a} was fabricated using two buried gates. The top gate is located very close to the graphene layer and produces a very sharp junction. Conversely, the buried gates are located under a thick oxide layer and have a large spacing between them. This configuration of buried gates yields a long junction width. 

In Fig. (\ref{dw_dependence}) we demonstrate the dependence of Landau level mixing on junction width and applied magnetic field. We simulate the same configuration as in Fig. (\ref{transmission}), but this time we fix the device as a symmetric n-p junction with $\nu_n=-\nu_p=6$ and vary the magnetic field and junction width. Each point is an ensemble average of 400 simulations with different realizations of disorder, with a fixed interface roughness profile. 

For small junction widths, less than 40 nm, the simulations display full mixing of the Landau levels. For long junction widths, approximately 100 nm and greater, only the lowest Landau level mixes. In between the full mixing and and lowest level mixing regime, the conductance smoothly decreases with junction width.

Furthermore, the applied magnetic field can control the degree of Landau level mixing. We observe that for an applied magnetic field of 14 T, it is significantly easier to inhibit mixing at the junction. This is due to increased confinement of the Landau levels at higher magnetic fields, which reduces the junction width required to fully separate the Landau levels.

\subsection{Landau level mapping}
%%%%%%%%%%%%%%%%%%%%%%%%%%%%%%%%%%%
%%%%%%%    Disordered LLS   %%%%%%%
%%%%%%%%%%%%%%%%%%%%%%%%%%%%%%%%%%%
\begin{figure}
\includegraphics[width=0.5\textwidth]{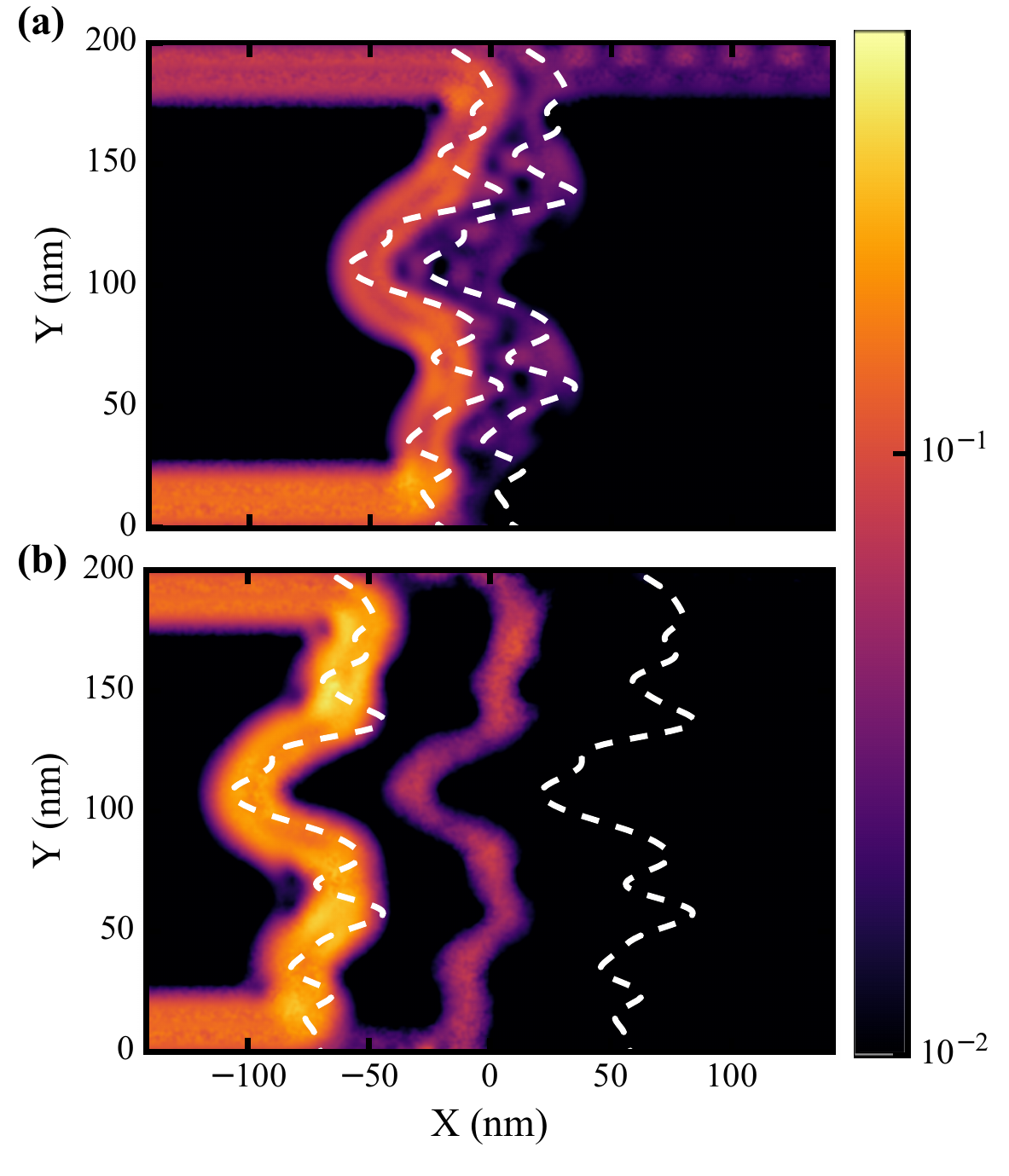}
\caption{Maps of the non-equilibrium carrier density injected from the left contact for two different junction widths, \textbf{(a)} 30 nm and \textbf{(b)} 125 nm. The full model described in the text is included; there is a finite junction width, interface roughness, and delta function disorder in the transition between the n-region (left side) and the p-region (right side). The edges of the junction are indicated by white dashed lines. The electron density enters from the bottom left of the junction, travels along the edge, turns up at the junction interface, and then turns either left or right at the top of the junction. In this case we configure the junction the same as in Fig. (\ref{dw_dependence}), with $E_1=-E_2=0.0875$ eV and $B=4$ T. In \textbf{(a)}, the two Landau levels are essentially located on top of each other. This is an example of a junction which would mix Landau levels and show plateaus in the ambipolar conductance. In \textbf{(b)}, the $1^{\textrm{st}}$ Landau level turns up at the junction interface, while the $0^{\textrm{th}}$ Landau level continues until it turns at the middle of the junction. At a junction width of 125 nm, the Landau levels are spatially separated and the device will only show a single plateau from the mixing of the $0^{\textrm{th}}$ Landau level.}
\label{disordered_LLs}
\end{figure}
%%%%%%%%%%%%%%%%%%%%%%%%%%%%%%%%%%%
%%%%%%%    Disordered LLS   %%%%%%%
%%%%%%%%%%%%%%%%%%%%%%%%%%%%%%%%%%%
%%%%%%%%%%%%%%%%%%%%%%%%%%%%%%%%%%%
%%%%%%%        LL pos       %%%%%%%
%%%%%%%%%%%%%%%%%%%%%%%%%%%%%%%%%%%
\begin{figure*}
\includegraphics[width=\textwidth]{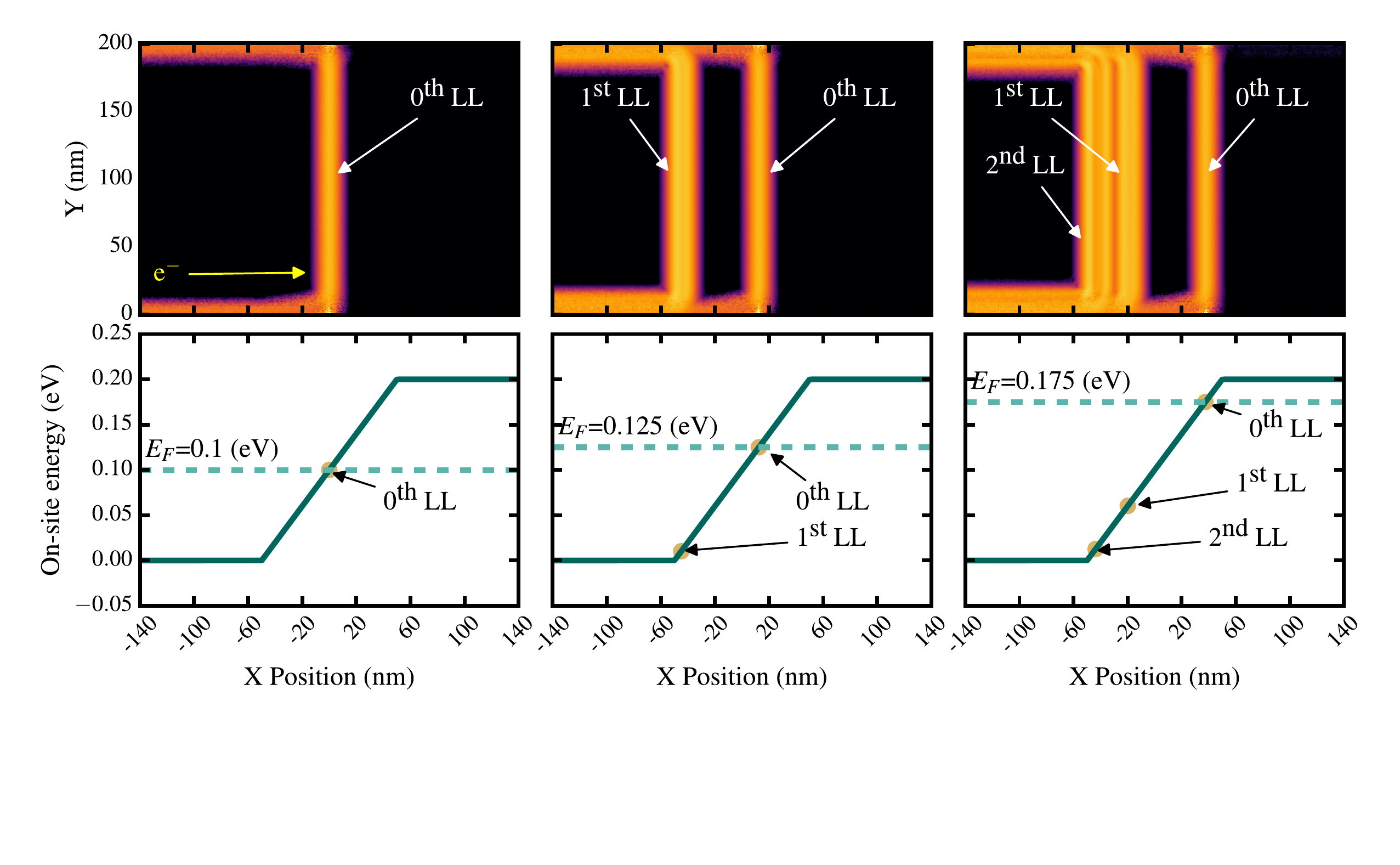}
\caption{Spatial maps of the electron probability density injected from the left contact. In these simulations we fix the on-site energy variation (solid black lines in the lower panels) and vary the Fermi energy (dotted lines in the lower panels). The junction width is 100 nm and the applied magnetic field is $B=10$ T. The position where each Landau level turns at the junction is closely predicted by (\ref{ELL_condition}). In \textbf{(a)}, $\nu_n=2$, only the $0^{\textrm{th}}$ Landau level transports. The current turns at the charge neutrality point. In \textbf{(b)},$\nu_n=6$, so two Landau levels transport. The $0^{\textrm{th}}$ Landau leve current turns at the charge neutrality point and the $1^{\textrm{st}}$ Landau level current turns close to the edge of the left junction interface. \textbf{(c)} Similar to \textbf{(b)}, but for $\nu_n=10$. The Landau level spacing in these plots is fixed by the slope of the on-site energy and the Landau level energies given by (\ref{energy_levels}). }
\label{pristine_LLs}
\end{figure*}
%%%%%%%%%%%%%%%%%%%%%%%%%%%%%%%%%%%
%%%%%%%        LL pos       %%%%%%%
%%%%%%%%%%%%%%%%%%%%%%%%%%%%%%%%%%%
%%%%%%%%%%%%%%%%%%%%%%%%%%%%%%%%%%%
%%%%%%%        LL Shape       %%%%%
%%%%%%%%%%%%%%%%%%%%%%%%%%%%%%%%%%%
\begin{figure}
\includegraphics[width=0.5\textwidth]{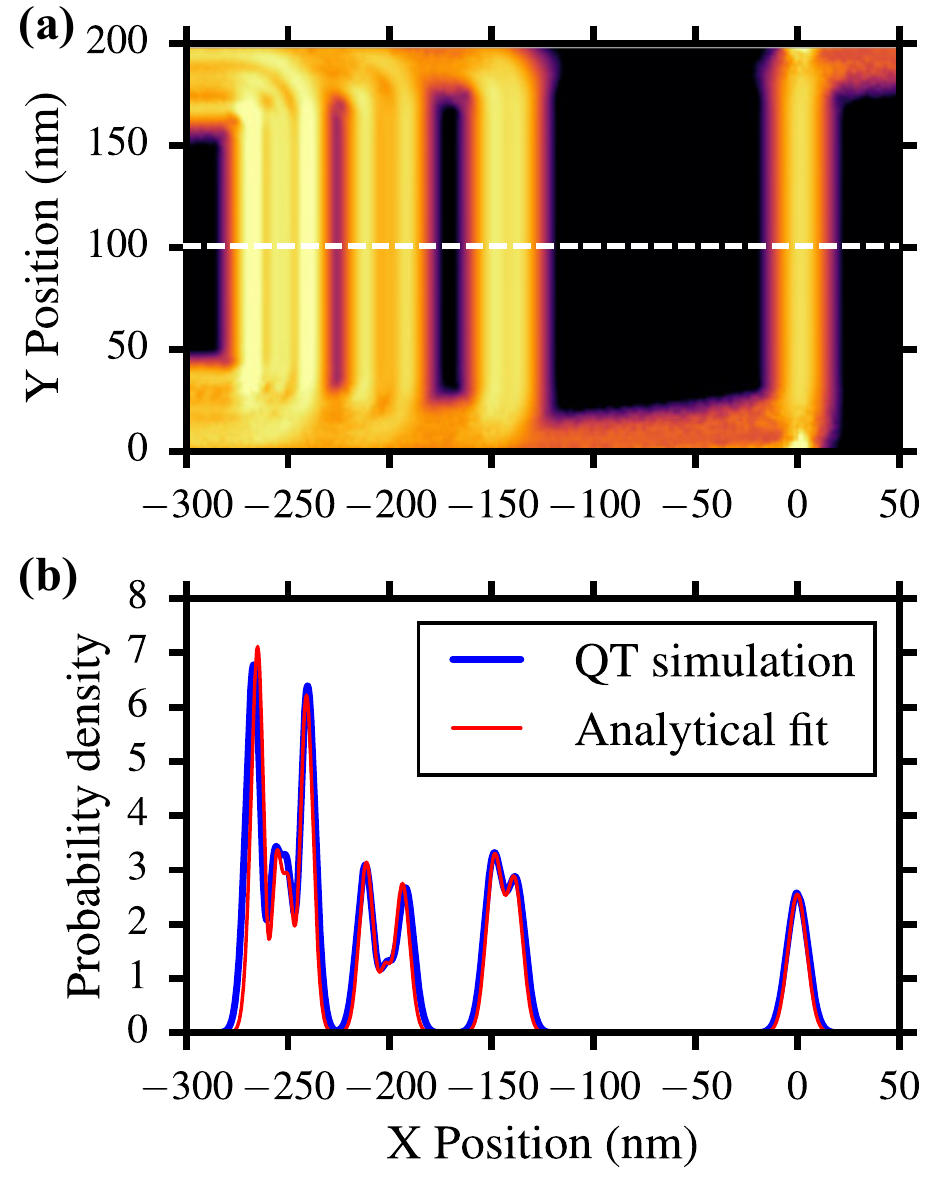}
\caption{\textbf{(a)} Spatial map of the electron probability density injected from the left contact. The applied magnetic field is $B$ = 4 T and $E_1 = - E_2 = 0.125$ eV, yielding a filling factor of $\nu_n=-\nu_p = 14$. We see four Landau levels transporting, each turning up at the junction at positions determined by (\ref{ELL_condition}). No roughness or delta function disorder is used, for clarity. Interestingly, the higher order Landau levels do not take on the form of Gaussians, instead having a more complex structure. In \textbf{(b)} we plot a slice of the simulation in \textbf{(a)} taken at $y =$ 100 nm (denoted by a white dashed line). In addition, we plot $P(x) = M_n\Psi_{n,k_y }(x)^{\dagger}\Psi_{n,k_y }(x) $, using the analytical form of the wave function (\ref{lukose_wf}) which was calculated in \cite{Lukose2007}.  }
\label{LL_shape}
\end{figure}
%%%%%%%%%%%%%%%%%%%%%%%%%%%%%%%%%%%
%%%%%%%        LL Shape       %%%%%
%%%%%%%%%%%%%%%%%%%%%%%%%%%%%%%%%%%

Here we will investigate the effect of junction width, $D_W$, on the distribution of Landau levels transporting across the junction. In Fig. (\ref{disordered_LLs}) we plot the non-equilibrium electron density injected from the left contact for $D_W=30 \textrm{ and } 100 \textrm{ nm}$. The device is configured as a symmetric n-p junction in a filling factor of $\nu_n=6$ and $\nu_p=-6$ on the left and right sides of the junction, respectively. The maps include the full model with a roughness profile given by (\ref{roughness_profile}) and the delta disorders inserted randomly between the dotted white lines. 

In Fig. (\ref{disordered_LLs}), the current travels along the bottom edge from the left contact, runs up the junction, and then turns either left or right at the top edge of the device. In the case of $D_W=30 \textrm{ nm}$, the $0^{\textrm{th}}$ and $1^{\textrm{st}}$ Landau levels mostly run on top of one another up the junction. There is a small separation in the levels, but much of the density overlaps. This map represents a junction which mixes higher order Landau levels and corresponds to a device which shows plateaus in the Hall conductance.

In the lower panel of Fig. (\ref{disordered_LLs}) we show a junction width of 125 nm. In this case, the $0^{\textrm{th}}$ and $1^{\textrm{st}}$ levels are fully separated at the junction. The $0^{\textrm{th}}$ level is located at the center of the junction, while the $1^{\textrm{st}}$ level is located at the left edge of the junction. This density map represents a device which only mixes the lowest Landau level, showing a single plateau in the ambipolar configuration. The spatial separation of Landau levels inhibits mixing at the junction, which is consistent with what was proposed by Klimov \textit{et al}\cite{NikolaiN.Klimov2015a}. 

Now that we have shown visualizations of the two classes of devices, those which mix many Landau levels and those which only mix the lowest Landau level, we will explore what determines the Landau level spacing. In Fig. (\ref{pristine_LLs}) we demonstrate a series of pristine density maps for the same potential profile, but with different Fermi levels. The three maps show a symmetric n-p junction ($\nu_n=-\nu_p = 2$), and two asymmetric n-p junctions ($\nu_n=6,10$ and $\nu_p=-2$). To examine the distribution of Landau levels more easily, we demonstrate these junctions without p-n interface roughness or delta disorder potentials.

The lower panels of Fig. (\ref{pristine_LLs}) show the energy band diagram of the junction, with the Fermi energy and corresponding Landau level energies. The position of the points for each Landau level is determined by the condition in (\ref{ELL_condition}). The condition in (\ref{ELL_condition}) also determines, within a few nanometers, where the current in the particular Landau level will turn at the junction. As the Fermi energy is increased in Fig. (\ref{pristine_LLs}), the $0^{\textrm{th}}$ Landau level moves from left to right, according to (\ref{ELL_condition}). When the energy of the first and second Landau levels is bigger than the on-site energy, $E_F - E_{LL1,2}>E_{\textrm{on-site}}(x)$, they begin to transport. The spacing between different Landau levels is fixed by the slope of the junction potential and follows the same parabolic form as given by (\ref{energy_levels}). We note that (\ref{ELL_condition}) works very well for predicting the turning point of the Landau levels, even without correcting the energy eigenvalues for electric field effects \cite{Lukose2007}. 

\subsection{Landau level shape}

Close examination of Fig. (\ref{pristine_LLs}) shows that higher order Landau levels split into multiple beams of carriers when they transport up the junction. The carriers do not simply travel at the junction as a Gaussian wave packet. The same effect may be seen in Fig. (\ref{disordered_LLs}), although it is slightly more difficult to see due to junction disorder. To investigate this effect, in Fig. (\ref{LL_shape}(a)) we show carrier density calculations for a symmetric n-p junction configured so that $\nu_n=-\nu_p = 14$. A junction width of 500 nm is simulated, which allows us to separately spatially resolve the first four Landau levels. Note that we zoom in to only show the left side of the junction, to more easily see the Landau levels.  

The higher order Landau levels shown in Fig. (\ref{LL_shape}) do not follow the simple Gaussian form of the $0^{\textrm{th}}$ Landau level, but instead split into a more complex structure. The shape of the Landau levels is influenced heavily by the A-B sub-lattice structure of graphene, where the Landau level is formed by the superposition of the wave functions from the A and B sub-lattices. 

In \cite{Lukose2007}, a closed form solution for the wave function of a graphene sheet with crossed electric and magnetic fields was obtained. The system studied by \cite{Lukose2007} is similar to ours, but the electric field in their work is pointed perpendicular to the edges of their graphene sheet. Despite this difference, their analytically calculated wave function may be compared to our simulated wave function when the junction width is very long. The wave function, adopted from \cite{Lukose2007}, is given by a two component spinor
\begin{equation}{\label{lukose_wf}}
\Psi_{n,k_y }(x,y) \propto e^{ik_y y}e^{-\frac{\theta}{2}\sigma_y}\begin{bmatrix}
sgn(n)\phi_{\mid n\mid-1}(\xi) \\
i\phi_{\mid n \mid}(\xi)
\end{bmatrix},
\end{equation} 
where 
\begin{equation}
\xi = \frac{(1-\beta^2)^{\frac{1}{4}}}{l_b}\left( x + l_b^2 k_y + sgn(n)\frac{\sqrt{2\mid n \mid}l_b\beta}{(1-\beta^2)^{\frac{1}{4}}} \right).
\end{equation}
The term $\beta = \frac{\mathcal{E}}{\nu_F B}$, where $\mathcal{E}$ is the applied electric field.

The wave function takes on the form of quantum harmonic oscillator functions, $\phi_{\mid n \mid}(\xi)$, where one sub-lattice contributes the $n^{\textrm{th}}$ harmonic oscillator function and the other sub-lattice is $n-1$. The index $n$ is an integer equal to the particular Landau level number. The $0^{\textrm{th}}$ Landau level is contributed by one sub-lattice and is a Gaussian. 

In Fig. (\ref{LL_shape}(b)) we show a slice of the simulated Landau level map at $y = 100 \textrm{ nm}$. We also calculate the probability density from (\ref{lukose_wf}), $P(x) = M_n\Psi_{n,k_y }(x)^{\dagger}\Psi_{n,k_y }(x) $, where $M_n$ is a normalization function used to match to the multi-moded transport in our simulation. We define the electric field for the analytical calculation as $\mathcal{E}=\frac{E_2-E_1}{eD_W}$. 

For the very long junction width considered in Fig. (\ref{LL_shape}), the wave function given by (\ref{lukose_wf}) may be applied to our simulations. The quantum transport simulation and analytical solution of \cite{Lukose2007} agree well for the $0^{\textrm{th}}$, $1^{\textrm{st}}$, and $2^{\textrm{nd}}$ Landau levels. The $3^{\textrm{rd}}$ Landau level straddles the edge of the junction where the electric field drops to zero and the uniform electric field assumption breaks down. Nevertheless, the analytical calculation still does a good job of describing the $3^{\textrm{rd}}$ Landau level as well. 

\section{Conclusions}

In this work we have studied the influence of junction width on Landau level mixing in ambipolar graphene p-n junctions. We utilized a combined p-n interface roughness and delta function disorder model, which represents a best case scenario to mix Landau levels. The model's capability to match experimental data on junctions which mix several Landau levels and those which only mix the lowest Landau level was demonstrated. Our simulations indicate that more disordered devices with short junction widths are likely to mix Landau levels, while cleaner devices with very wide junction widths will only mix the lowest Landau level. To support our arguments, we provided visualizations of non-equilibrium carrier density across the junctions and a demonstrated simple predictive model which determines how the Landau levels will separate at the junction. Finally, we compared our simulations with analytical calculations \cite{Lukose2007}, revealing the interesting form of higher order Landau levels. In the future, this model may be extended to more complex devices with multiple p-n junctions. \\

\begin{acknowledgments}
%\section*{Acknowledgments}
The authors acknowledge financial support provided by the U.S. Naval Research Laboratory (grant number N00173-14-1-G017). S. LaGasse acknowledges helpful discussion with T. Low regarding his interface roughness model. 
\\
\\
\\
\\
\\\\\\\\\\\\\\\
\\\\
\\
\\\\
\\
\end{acknowledgments} %without the \\, there is an anomalous spacing between the section header and text

%merlin.mbs apsrev4-1.bst 2010-07-25 4.21a (PWD, AO, DPC) hacked
%Control: key (0)
%Control: author (8) initials jnrlst
%Control: editor formatted (1) identically to author
%Control: production of article title (-1) disabled
%Control: page (0) single
%Control: year (1) truncated
%Control: production of eprint (0) enabled
%

%\bibliography{QHE_in_PNJ_manuscript}
\end{document}